\documentclass[3p,times]{elsarticle}

\usepackage{amsmath,amssymb,amsfonts}

\usepackage{graphicx}
\usepackage{textcomp}
\usepackage{graphicx}
\usepackage{epstopdf}
\usepackage{subfigure}
\usepackage{color}
\usepackage{float}
\usepackage{ntheorem}
\usepackage{ecrc}
\usepackage{ntheorem}
\newtheorem{rem}{Remark}
\volume{1}

\firstpage{1}

\journalname{Preprint submitted to Applied Mathematics and Computation}

\runauth{H. Ren, Y. Yan and F. Gao}

\jnltitlelogo{Version 2}

\usepackage{amssymb}
\usepackage[colorlinks,
            linkcolor=red,
            anchorcolor=blue,
            citecolor=green
            ]{hyperref}
\def\sectionautorefname~#1\null{Section~#1\null}
\def\subsectionautorefname~#1\null{Section~#1\null}
\def\subsubsectionautorefname~#1\null{Section~#1\null}
\def\figureautorefname~#1\null{Fig.~#1\null}
\def\tableautorefname~#1\null{Tab.~#1\null}
\def\equationautorefname~#1\null{Eq.~(#1)\null}
\newcommand{\refs}[1]{%
  \begingroup%
  \def\figureautorefname~##1\null{Figs.~##1\null}%
  \def\equationautorefname~##1\null{Eqs.~(##1)\null}%
  \autoref{#1}%
  \endgroup%
}

\usepackage[figure]{hypcap}
\biboptions{sort&compress}
\usepackage[figuresright]{rotating}

\begin{document}
\begin{frontmatter}



\dochead{}

\title{Variable guiding strategies in multi-exits evacuation: Pursuing balanced pedestrian densities}


\author[Ren and Gao]{Huan Ren}
\author[Yan]{Yuyue Yan\corref{mycorrespondingauthor}}\cortext[mycorrespondingauthor]{Corresponding author.}\ead{yan.y.ac@m.titech.ac.jp}    % Add the
\author[Ren and Gao]{Fengqiang Gao}            % e-mail address
  % (ead) as shown
\address[Ren and Gao]{School of Information Science Technology, Xiamen University Tan Kah Kee College, Fujian 363105, China}
\address[Yan]{Department of Systems and Control Engineering, Tokyo Institute of Technology, Tokyo 152-8552, Japan}  % Please supply

\begin{abstract}
Evacuation assistants and their guiding strategies play an important role in the multi-exits pedestrian evacuation. To investigate the effect of guiding strategies on evacuation efficiency, we propose
a force-driven cellular automaton model with adjustable guiding
attractions imposed by the evacuation assistants located in the exits. In this model, each of the evacuation assistants
tries to attract the pedestrians in the evacuation space towards its own exit by sending a quantifiable guiding signal,
which may be adjusted according to the values of pedestrian density near the exit.
The effects of guiding strategies pursuing balanced pedestrian densities are studied.
It is observed that the unbalanced pedestrian distribution is mainly yielded by a snowballing effect
generated from the mutual attractions among the pedestrians, and can be suppressed by controlling the pedestrian densities around the exits.
We also reveal an interesting fact that given a moderate target density value, the density control for the partial regions (near the exits)
could yield a global effect for balancing the pedestrians in the rest of the regions and hence improve the evacuation efficiency.
Our findings may contribute to give new insight into designing effective guiding strategies in the realistic evacuation process.
\end{abstract}


\end{frontmatter}

% \linenumbers

\section{Introduction}
The efficiency estimation for pedestrian evacuation is a key aspect of  safety performance evaluation when the earthquake, fire, terrorist attacks, and other emergencies occur in public places. During the evacuation process, pedestrians may possess some irrational behaviors such as anxiety, panic and blindly following, to name but a few, which may affect the evacuation efficiency and hence cause some casualties and property losses \cite{meng2019pedestrian,cheng2018emergence,haghani2019panic,guan2019towards}.
To avoid such cases, it is significantly important to understand the pedestrian behaviors in the evacuation process \cite{lovaas1994modeling,helbing2000simulating,muller2014study}.
In the literature, the existing pedestrian dynamics are mainly categorized to the continuous-time and discrete-time models \cite{PhysRevE,varas2007cellular,huang2017behavior,alizadeh2011dynamic,cao2018exit,shahhoseini2019pedestrian,huang2015behavioral,chraibi2010generalized,liao2014layout,hao2014exit,HRABAK2017486}.
As the typical discrete-time dynamics, the cellular automaton models provide a structure for reflecting the pedestrians' microscopic characteristics
with some simple update laws (evolution rules), which may be combined with the notion of social force \cite{chen2012study}, floor field \cite{varas2007cellular,li2019extended}, etc.
{For instant, Huang et al. established a cellular automaton model to describe the movement of heterogeneous pedestrians in terms of cooperative behavior and
examined how the dynamics of dependency relationship among the pedestrians affect the evacuation efficiency  \cite{huang2017weighted}.
%Alizadeh investigated a dynamic cellular automaton model for a evacuation space with obstacles and discussed the influences of doors position and width, pedestrians distribution on the evacuation efficiency \cite{alizadeh2011dynamic}.
Pereira et al. extend the  cellular automaton model to allow  pedestrians to change direction for accessing an alternate exit route \cite{pereira2017emergency}. Miyagawa and Ichinose proposed a multi-grid cellular automaton model with turning behavior \cite{miyagawa2020cellular}.}

During the emergency evacuation, some evacuation assistants are usually adopted to help evacuees escape as soon as possible. The related works in terms of guided behaviors in evacuation can be found in  \cite{yang2016necessity,ma2016effective,ma2017dual,zhou2019guided,zhou2019optimization,long2020simulation,yang2020guide}. In those works, most of the scholars only discussed the single-exit evacuation scenario based on a social force model.
For example, Yang et al. \cite{yang2016necessity} first claimed the necessity of guides in pedestrian emergency evacuation.
Ma et al. \cite{ma2016effective,ma2017dual} considered dynamic leaders inside a room who attract the crowds and move together with them towards a unique exit with different  pedestrian densities.
It was found in \cite{ma2016effective} that
a large evacuees crowd can be efficiently guided by those dynamic evacuation assistants if the neighbor density is moderate.
{However, the possibility and effects of controlling the pedestrian densities are not considered yet.}

{Different from the existing works, in this paper, we consider the situation where multiple evacuation assistants locate in their corresponding exits and may try to adjust their guiding signals to attract the crowds under density control.} It is well known that the balanced pedestrians evacuation could improve the evacuation efficiency \cite{kurdi2020balanced}.
Hence, we mainly focus on pursuing balanced pedestrian densities in a cellular automaton-based evacuation model with multiple exits, and discuss the inherent relation between the evacuation efficiency and the balanced pedestrian densities.  Specifically, we suppose that each of the evacuation assistants measures the pedestrian density around its own exit and changes the guiding strategies (i.e., the strength of the guiding signal) according to the measured data.

The contributions of this paper are summarized as follows.  1) We propose a novel force driven cellular automaton model with multiple exits and variable guiding strategies, where the strength of the guiding signal for each of the exits (evacuation assistants) is adjustable according the evacuation situation.
2) The reason why unbalanced pedestrian distribution yielded in the evacuation process is interpreted.
3) Two different density control laws for obtaining the on-off and quantifiable guiding signals are characterized for suppressing the unbalanced pedestrian distribution. The relation between the target density and the evacuation efficiency for each of the proposed control laws is discussed and the optimal target density {is numerically found for a square evacuation space}.

%\cite{ma2016effective,ma2017dual,li2017driving,guan2020cooperative,li2020relationship,cao2018exit,huang2017behavior,zhou2018optimal,zhou2019cellular,kurdi2020balanced,zhou2019guided,zhou2019optimization,li2019extended,yang2020guide,long2020simulation,ZHANG2021488}\cite{yue2011simulation2}.

The rest of this paper is organized as follows: A novel force driven cellular automaton model with multiple exits and variable guiding strategies as well as the two designed density control laws are proposed in Section~\ref{section2}. In Section~\ref{section:3}, we first interpret the fact that the guiding behaviors without density control may waste the capacity of the exits, and then we apply the density control laws designed in Section~\ref{section2} to the evacuation model to observe the effects of the control behaviors for the evacuation process. The comparisons of the evacuation efficiencies with and without density control are also given in  Section~\ref{section:3}. Finally, we conclude our paper in Section~\ref{section:4}.

\emph{Notation:} In this paper, we write $\mathbb R_+$ for the set of positive real numbers, $\mathbb N_+$ for the set of positive integers and $\emptyset$ for the empty set.

\vspace{-10pt}
\section{Model}\label{section2}\vspace{-4pt}
In this section, we propose our pedestrian evacuation model based on force-driven cellular automaton \cite{chen2012study} with \emph{time-varying} guiding signals. In this model, the state space (i.e., evacuation space) is divided to $N\times N$ number of cells where each of the cells may be occupied by a single pedestrian. At each time instant (or, equivalently, time step), the pedestrians may have multiple desired moving directions and need to compete with the others if the desired cell is targeted by more than one pedestrians. For each of the pedestrians, we assume that there are eight possible moving directions whose corresponding cells neighbor to the current location (cell). Defining a social (resultant) force consisting of a guiding force from an evacuation assistant, interaction forces among pedestrians and an attractive force from a visible exit, the moving direction candidates for each of the pedestrians can be properly given according to the directions of the decomposed forces at $x$, $y$, $s$ and $f$-axis illustrated in \autoref{fig:2} below, where the preferences of those moving direction candidates solely depend on the magnitudes of the decomposed forces. In the following statements, we first define the component forces for the pedestrians to describe their self-organization behaviors and then present the evolution rules for the proposed model.
\begin{figure}
  \centering
  \includegraphics[width=65mm]{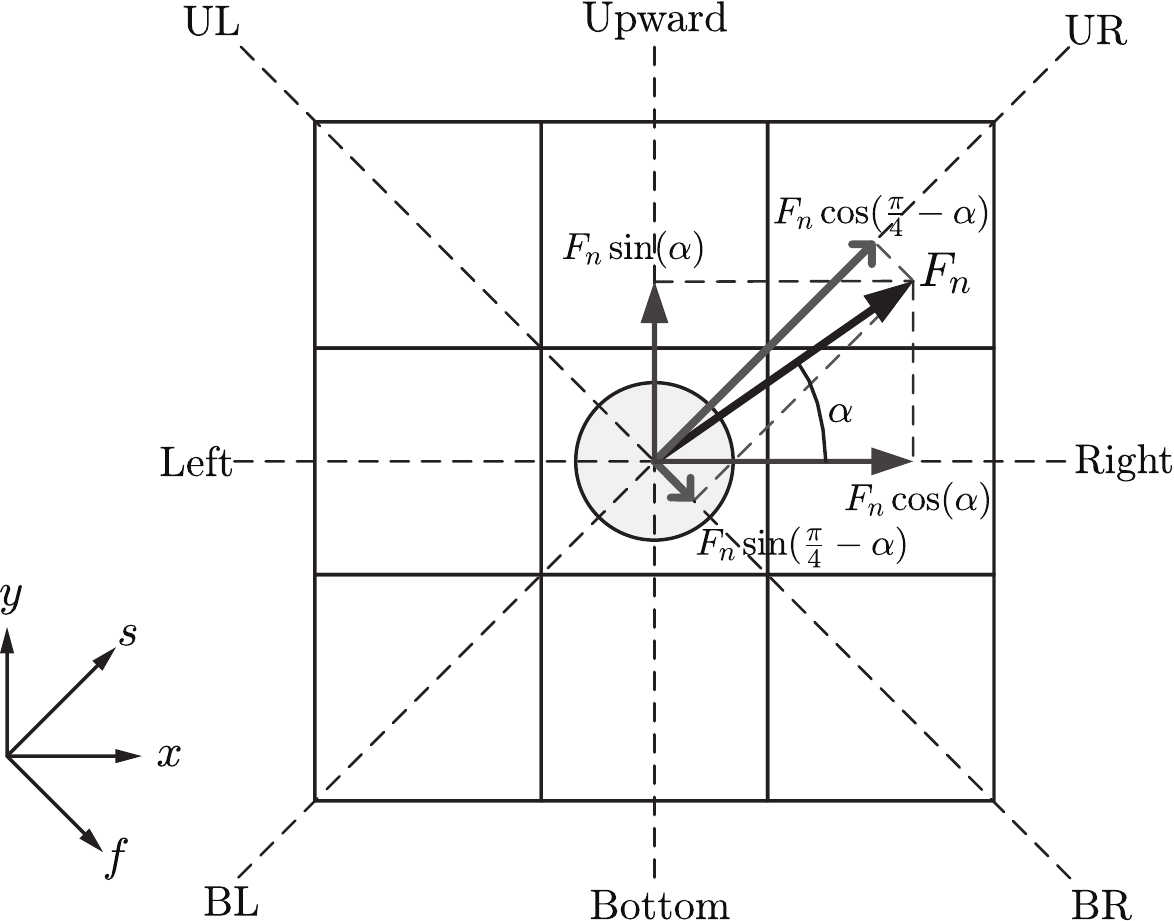}\vspace{-8pt}
  \caption{Decomposed forces in \emph{x}, \emph{y}, \emph{s} and \emph{f} axis. The ranking on the magnitudes of decomposed forces is given by upper right, right, upward and bottom right and hence this pedestrian is assumed to prefer those four directions as the next moving directions. If the preferred moving direction is in a conflict with the other pedestrian, then a competition is considered to determine the winner who occupy the competed cell in the next time instant (time step).   If the pedestrian lost in all of the competitions for the four preferred moving directions, then he remains in the original cell in the next time instant.   }
  \label{fig:2}\vspace{-8pt}
\end{figure}

\vspace{-5pt}
\subsection{Component Forces} \label{Sec22}
Denoting $\mathbb {RM}(k)$ as the set of remaining pedestrians in the state space at the time instant $k=0,1,2,\ldots$, we define the component social forces for the pedestrian $n\in\mathbb {RM}(k)$ in the following statements.
\vspace{-5pt}\subsubsection{Guiding Force}
In reality, the pedestrians may receive guiding signals (i.e., instructions) from the evacuation assistant(s), e.g., sound, action and warning lights, and hence the moving directions and evacuation efficiency are significantly influenced. In this paper, we assume that each of the exits possesses an independent evacuation assistant who attract the pedestrians $n\in\mathbb {RM}(k)$ towards its own exit. Let $\mathcal N\triangleq \{1,\ldots,N_e\}$ denote the set of exits (or, evacuation assistant) where $N_e$ denotes the number of exits.
For each of the pedestrians $n\in\mathbb {RM}(k)$, the guiding force is defined by
{\setlength\abovedisplayskip{3pt}
\setlength\belowdisplayskip{3pt}
\begin{equation}\label{guiding}
    F_{\rm guide}^n = u_{I_n}D  \vec r_{n,I_n},
\end{equation}
where} $I_n\in\mathcal N$ denotes the index of the evacuation assistant believed by the pedestrian $n$, $\vec{r}_{n,I_n}$ denotes a unit vector from the pedestrian $n$ to the exit $I_n$, $D\in\mathbb R _+$ denotes the maximal guiding intensity, and $u_{I_n}\in[0,1]\subset\mathbb R$, $I_n\in\mathcal N$, denote coefficients designed (or controlled) by the evacuation assistants $I_n\in\mathcal N$. If $u_i$ is taken as a zero (resp., nonzero) value, then it is understood that the evacuation assistant $i\in\mathcal N$ stops attracting (resp., tries to attract) the pedestrians $n\in\mathbb {RM}(k)$.
Inspired by the notion of \emph{signal to interference ratio}, we consider{\setlength\abovedisplayskip{3pt}
\setlength\belowdisplayskip{3pt}
\begin{equation}
   I_n\triangleq\arg\max_{i\in\mathcal N}\left(\frac{u_i}{1+\sum_{j\in\mathcal N\setminus\{i\}}(r_{n,i}^2/r_{n,j}^2)}\right),\label{k}
\end{equation}
with} $r_{n,i}\in\mathbb R_+$ denoting the distance (step length) from the pedestrian $n$ to the exit $i$, $i\in\mathcal N$.
Note that the expression in the right-hand side of \autoref{k} represents the reliability of the guiding signal from the evacuation assistant $i\in\mathcal N$. Specifically, if $u_i=1$, $i\in\mathcal N$, then the pedestrian $n$ is attracted by the nearest evacuation assistant. However, if the nearest evacuation assistant $i\in\mathcal N$ turns off its guiding signal, i.e., $u_i=0$, then the pedestrian $n$ is attracted by the other evacuation assistants.
Hence, we note that the coefficients $u_i$, $i\in\mathcal N$, may be time-varying according to the strategies of the evacuation assistants (see the controlled coefficients $u_i$, $i\in\mathcal N$, in \autoref{controller} below).

%\begin{figure}
%  \centering
%  \includegraphics[width=20mm]{Fig4.eps}\vspace{-8pt}
%  \caption{Delete it later.}
%  \label{fig:4}\vspace{-8pt}
%\end{figure}
\vspace{-5pt}
\subsubsection{Mutual forces among pedestrians}
The existence of mutual forces among the pedestrians is the main reason of herding effect. Let $\mathbb {VP}_n$ denote the set of visible pedestrians for the pedestrian $n\in\mathbb{RM}(k)$, i.e.,
the distance $r_{nm}\in\mathbb N_+$ from the pedestrian $n$ to $m$ is smaller than a constant integer $w\in\mathbb N_+$ for all $m\in\mathbb{VP}$.
According to the distance $r_{nm}$, we define the mutual force{\setlength\abovedisplayskip{3pt}
\setlength\belowdisplayskip{3pt}
\begin{eqnarray}\label{mutual}
F_{\rm mutual}^{n,m}= \begin{cases} -{\eta_1}\vec{r}_{n,m},& \mbox{if $r_{n,m}=1$} \\
  {\eta_2}/{r_{n,m}^2}\vec{r}_{n,m}, &\mbox{if $1<r_{n,m}\leq w$}\end{cases}, \quad m\in \mathbb {VP}_n,
 \end{eqnarray}
where} $\eta_1\in\mathbb R_+$ denotes a coefficient for repulsion, $\eta_2\in\mathbb R_+$ denotes a coefficient for attraction, $\vec{r}_{n,m}$ denotes a unit vector from the pedestrian $n$ to the visible pedestrian $m$, and $w$ denotes the length of the filed of view.
%Therefore, the interaction forces consider the cases of attraction and repulsion and inefficacy. Taking reference of the coulomb force, the pedestrian in different distance are assumed to be the like charges or unlike c harges, and the quantity of electricity is asserted to be 1.0 Unit.$^{12,14}$  Thus,
\vspace{-5pt}
\subsubsection{Attractive force to a visible exit}
If one of the exits appears in the visible region, the pedestrian $ n\in\mathbb {RM}(k)$ is supposed to possess an attractive force to the visible exit, which is defined by{\setlength\abovedisplayskip{3pt}
\setlength\belowdisplayskip{3pt}
\begin{equation}\label{exit}
    F_{\rm exit}^{n,i}=  E\vec{r}_{n,i}, \quad i\in\mathbb {VE}_n\triangleq\{i\in\mathcal N:  r_{n,i}\leq w\},
\end{equation}
where} $E\in\mathbb R_+$ denotes a constant value, $r_{n,i}$ and $\vec{r}_{n,i}$ denote the distance and the unit
vector from the pedestrian $n$ to the exit $i\in\mathcal N$, respectively.
\vspace{-5pt}
\subsection{Resultant force and preference moving direction}
Consequently, for a pedestrian $ n\in\mathbb {RM}(k)$, the resultant social force is defined by{\setlength\abovedisplayskip{3pt}
\setlength\belowdisplayskip{3pt}
\begin{equation}\label{social}
    {F_{n} =    w_1 F_{\rm guide}^n+w_2\!\sum\nolimits_{i\in\mathbb {VE}} F_{\rm exit}^{n,i}+w_3\!\sum\nolimits_{m\in\mathbb {VP}}{F_{\rm mutual}^{n,m}},}
\end{equation}
where} {$w_1$, $w_2$ and $w_3\in\mathbb R_+$ denote the positive weighting factors depending on the evacuation scenario. For instant, the influence of guiding signal and visible exit may be \emph{weakened} (i.e., $w_1<w_3$ and $w_2<w_3$ hold) when the pedestrians are trapped by panic as the space is full of the smoke of fire.}
We assume that the preferences of moving direction
candidates for the pedestrian $ n\in\mathbb {RM}(k)$ are determined by sorting the magnitude of the decomposed forces of \emph{F}$_{n}$ at \emph{x}, \emph{y}, \emph{s} and \emph{f}-axis (see \autoref{fig:2}). For example, denoting the cell address at the upward, right, upper right and bottom right directions by $Q_{\rm Upward}^n$, $Q_{\rm Right}^n$, $Q_{\rm UP}^n$, and $Q_{\rm BR}^n$, the ordered preference target cells
 in \autoref{fig:2} are given by
[$Q_{\rm UR}^n$, $Q_{\rm Right}^n$, $Q_{\rm Upward}^n$, $Q_{\rm BR}^n$]. {However, if one of the preferred target cells is already occupied, then the corresponding decomposed force is understood as zero and hence the sequence of ordered preference target cells may be altered.}  For the following statements, we define $P_n=[P_n^1,P_n^2,P_n^3,P_n^4]$ as the sequence of ordered preference target cells where $P_n^i$ denotes the address of the $i$-th preferred target.
{Here, it is important to note that even though the preference target cells are deterministic, there may still exist some \emph{stochastic} processes among deciding addresses of pedestrians at the next time instant, which are described as \emph{random competitions} shown in \autoref{evolution}~ii) below.}

\vspace{-5pt}
\subsection{Evolution rules}\label{evolution}
Given an initial set $\mathbb {RM}(k)$ with $k=0$, the evolution rules are summarized as follows.

i) Each of the pedestrians $ n\in\mathbb {RM}(k)$ remaining in the state space calculates his own resultant social force according to \refs{guiding}--(\ref{social}),
and determines his ordered preference target cells (unoccupied) as moving direction candidates according to the magnitude of the decomposed forces.

ii) Target competition. {Recalling that multiple pedestrians may have the same preferred target cells, which is so called `collision effect' in  \cite{tanimoto2010study} and \cite{tanimoto2019evolutionary}, we consider target competitions among the pedestrians to solve the collisions. In particular,
we assume that each pedestrian $n$ conflicting with the others possesses at most 4 rounds of chances to compete for the preferred target cell according to the sequence $P_n$.
If the pedestrian $n$ wins in the competition at the round $i=1,2,3,4$, then his $i$-th preferred target cell (i.e., $P_n^i$) is assigned to him as the address at next time instant and the others compete for the next un-assigned preferred target cell. If the pedestrian loses in all the competitions, then he stays in his current location at next time instant because of the \emph{unsolvable collisions}. Here, we assume the winner is \emph{randomly} selected in the competitions.}

iii) Update the addresses for all the pedestrians and delete the pedestrians who reach one of the exits. Let $k=k+1$ and update the set $\mathbb {RM}(k)$.

iv) If all of pedestrians have left the state space, i.e., $\mathbb {RM}=\emptyset$, the program ends. Otherwise, go to i) to continue.
\vspace{-3pt}
\begin{rem}
\emph{{Note that depending on the parameters defined in \autoref{Sec22} and \autoref{social}, our proposed model can reflect \emph{various} evacuation situations such as emergent evacuations in face of potential risk, and the ones considering panic evacuees, illegible exits and time-varying guiding signals. For example, supposing that $w_2$ in \autoref{social} is very small but $\eta_2$ in \autoref{mutual} is very large, the proposed model can reflect the scenario where evacuees are exposed to critical situation such as fires. In such a case, the pedestrians may hardly find the exits in their visions, be
trapped by panic, and never leave the evacuation space in the simulation. } }
\end{rem}\vspace{-6pt}

\vspace{-5pt}
\subsection{Strategy on coefficients $u_i$, $i\in\mathcal N$}\label{controller}
In the following sections, we investigate the strategies to change the coefficients $u_i$, $i\in\mathcal N$, for adjusting the guiding strength of the evacuation assistants.
Specifically, we
consider pedestrian density control in the evacuation and characterize its influence to
the evacuation efficiency. Let $\rho_i$ denote the pedestrian density near the exit $i\in\mathcal N$, which is sampled by the corresponding evacuation assistant.
We utilize two schemes to update the values of $u_i$, $i\in\mathcal N$.

1) Scheme 1 (Bang-bang control): Let $\rho_{\rm aim}$ be a target density. The values of $u_i$, $i\in\mathcal N$, are determined by {\setlength\abovedisplayskip{3pt}
\setlength\belowdisplayskip{3pt}
\begin{eqnarray}\label{BB}
u_i(k)=\left\{\begin{array}{c}
                        1,\quad  \rho_i(k-1)\leq\rho_{\rm aim},\\
                        0,\quad  \rho_i(k-1)>\rho_{\rm aim},
                      \end{array}
\right. \quad i\in\mathcal N,\quad k=1,2,3,\ldots,
\end{eqnarray}
where} the guiding signal is off (resp., on) when the current density is lager (resp., smaller) than the target density.

2) Scheme 2 (PI control): Suppose $u_i$, $i\in\mathcal N$, are quantifiable. Let $\rho_{\rm aim}$ be a target density and let $e(k)=\rho_i(k)-\rho_{\rm aim}$.  The values of $u_i$, $i\in\mathcal N$, are determined by the following update laws \vspace{-1pt}{\setlength\abovedisplayskip{2pt}
\setlength\belowdisplayskip{2pt}
\begin{equation}\label{8}
u_i(k)=u_i(k-1)+K_pe(k)+K_i\sum\nolimits_{j=0}^ke(j), \quad i\in\mathcal N,\quad k=1,2,3,\ldots,
\end{equation}
where} ${K_p}$ an ${K_i}$ are some positive constants.
Under the above update law, the guiding strength is reduced (resp., increased) when the current density is lager (resp., smaller) than the target density.

\vspace{-9pt}
\section{Simulation and Results}\label{section:3}\vspace{-4pt}
\begin{figure}
  \centering
  \includegraphics[width=110mm]{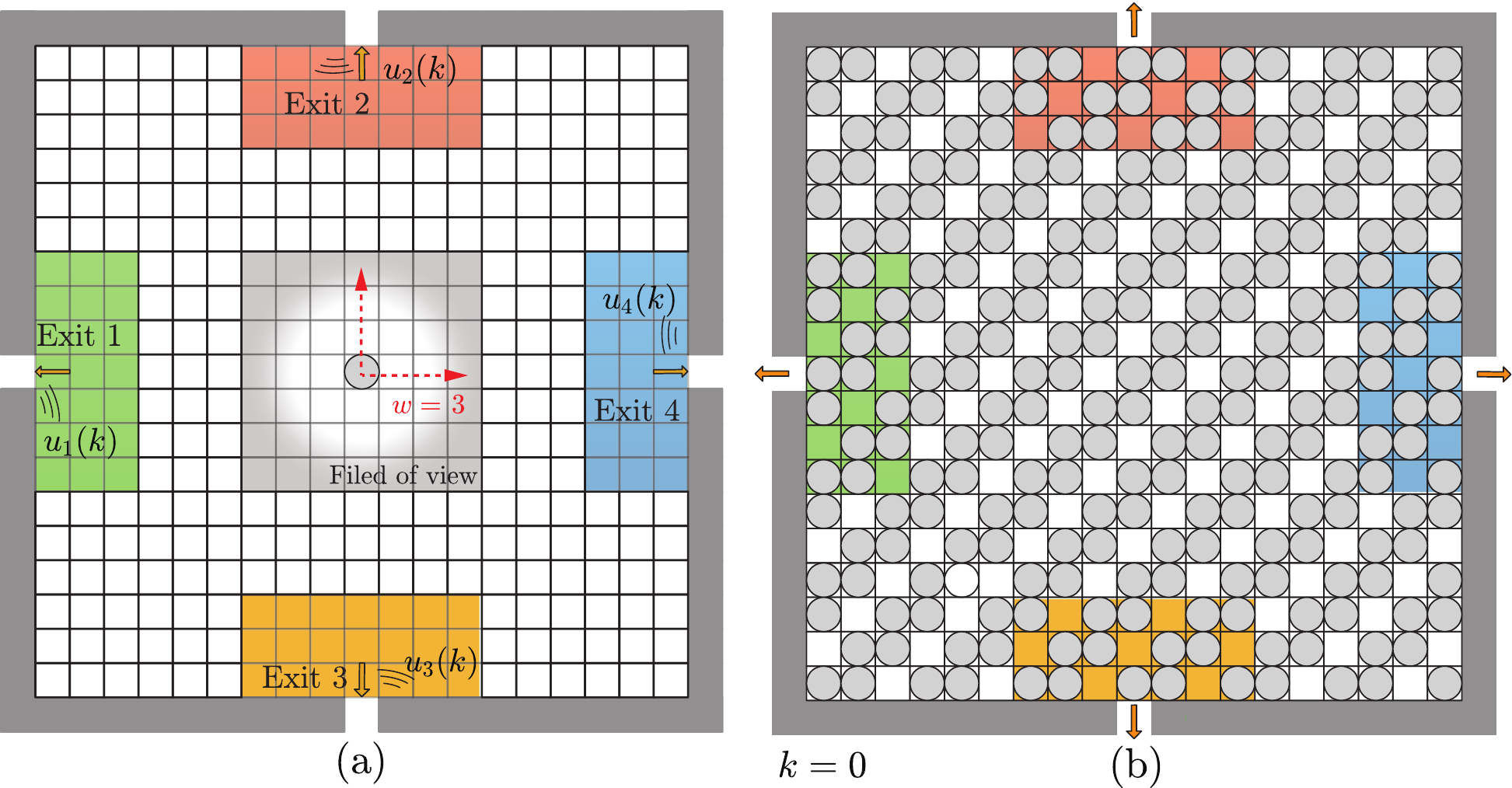}\vspace{-10pt}
  \caption{Evacuation space (i.e., state space) and initial distribution in simulations. (a): Simulation environment with $N=19$ and $\mathcal N=\{1,2,3,4\}$, where the colored areas near exits denote the regions used for observing (or, calculating) pedestrian densities, (b): initial distribution of pedestrians ($\mathbb {RM}(0)$ includes 241 elements) where circles denote the pedestrians.}
  \label{fig:5}\vspace{-11pt}
\end{figure}
In this section, we present the simulation results of the proposed model under the different strategies in terms of $u_i(k)$, $i\in\mathcal N$. We consider a square room with $N_e=4$ exits as the evacuation environment, i.e., $\mathcal N=\{1,2,3,4\}$. We divide the evacuation environment to $19\times 19$ cells and establish the state space as illustrated in \autoref{fig:5}(a), where the length of the filed of view is set to 3 (i.e., $w=3$) and the 4 regions for observing pedestrian densities are given. The initial amount of pedestrians is set to 241. Pedestrians at the time instant $k=0$ are distributed as \autoref{fig:5}(b). Since each of the 4 regions near the exits only possesses 21 cells, the pedestrian density near the exit $i\in\mathcal N$ is measured as $\rho_i(k)={\rm num}_i(k)/21$, where ${\rm num}_i$ denotes the number of pedestrians located in the region $i$. Hence it can be seen from \autoref{fig:5}(b) that the initial pedestrian densities in all of the 4 regions are the same as each other and given by $\rho_i(0)=0.667$, $i\in\mathcal N$.
The maximal guiding intensity is set to 30 (i.e., $D=30$ in \autoref{guiding}), the parameter for attractive force to a visible exit is set to 40 (i.e., $E=40$ in \autoref{exit}), and the coefficient for mutual repulsion (resp., attraction) among the pedestrians is set to 0.6 (resp., 1.2), i.e., $\eta_1=0.6$ and $\eta_2=1.2$ in \autoref{mutual}. {The weighting factors in \autoref{social} are set to $w_1=w_2=w_3=1$.}
\vspace{-6pt}
\subsection{Compared Parameters}
We define the following two notions to compare the evacuation efficiency.

A) Unbalanced degrees in terms of density: $\alpha_i$, $i\in\mathcal N$

Unbalanced factor $\alpha_i$ represents how much the pedestrian density near the exit $i\in\mathcal N$ is unbalanced comparing to the ones for the other exits, and is given by {\setlength\abovedisplayskip{2pt}
\setlength\belowdisplayskip{2pt}
\begin{equation}
  \alpha_i(k)=\left|\frac {\rho_i(k)}{\sum_{j\in \mathcal N}\rho_j(k)}-\frac {c_i}{\sum_{j\in \mathcal N}c_j}\right|,\quad i\in\mathcal N, \quad k=1,2,3,\ldots,
\end{equation}
where} \emph{$\rho$}$_{i}$ denotes the pedestrian density near the exit \emph{i} (see the colored region in \autoref{fig:5}(a)), $i\in\mathcal N$, and $c_{i}$ denotes a constant value representing the capacity of exit \emph{i}. Since all the exits considered in \autoref{fig:5} have the same size and hence same capacity, we have $\alpha_i(k)=\Big|\frac {\rho_i(k)}{\sum_{j\in \mathcal N}\rho_j(k)}-\frac {1}{N_e}\Big|\in\Big[0,\frac {N_e-1}{N_e}\Big]$.

%B) Unbalanced degrees in terms of handled amount: $\beta_k$, $k\in\mathcal N$
%
%Unbalanced factor $\beta_k$ represents how much the amount of pedestrians moving through the exit $k$ is unbalanced comparing to the ones for the other exits, and is given by
%\begin{equation}
%  \beta_k(k)=\left|\frac {n_k(k)}{\sum_{i\in \mathcal N}n_i(k)}-\frac {c_k}{\sum_{i\in \mathcal N}c_i}\right|=\left|\frac {n_k(k)}{\sum_{i\in \mathcal N}n_i(k)}-\frac {1}{N_e}\right|\in\Big[0,\frac {N_e-1}{N_e}\Big],
%\end{equation}
%where \emph{$n$}$_{k}$ denotes the amount of pedestrians moving through the exit $k$, $k\in\mathcal N$.

B) Travel time: $T_{\rm end}$

Travel time is the number of steps which the last pedestrian used to leave to the exit, i.e., $\mathbb {RM}(k)=0$ at $k=T_{\rm end}$.
Note that if the unbalanced degree maintain as a large number, then it is understood that the capacities of some exits are not efficiently utilized. In this case, $T_{\rm end}$ is probably larger than the case of balanced pedestrian density.

\subsection{Simulation Results Without Density Control}\label{without}

\begin{figure}
  \centering
  \includegraphics[width=165mm]{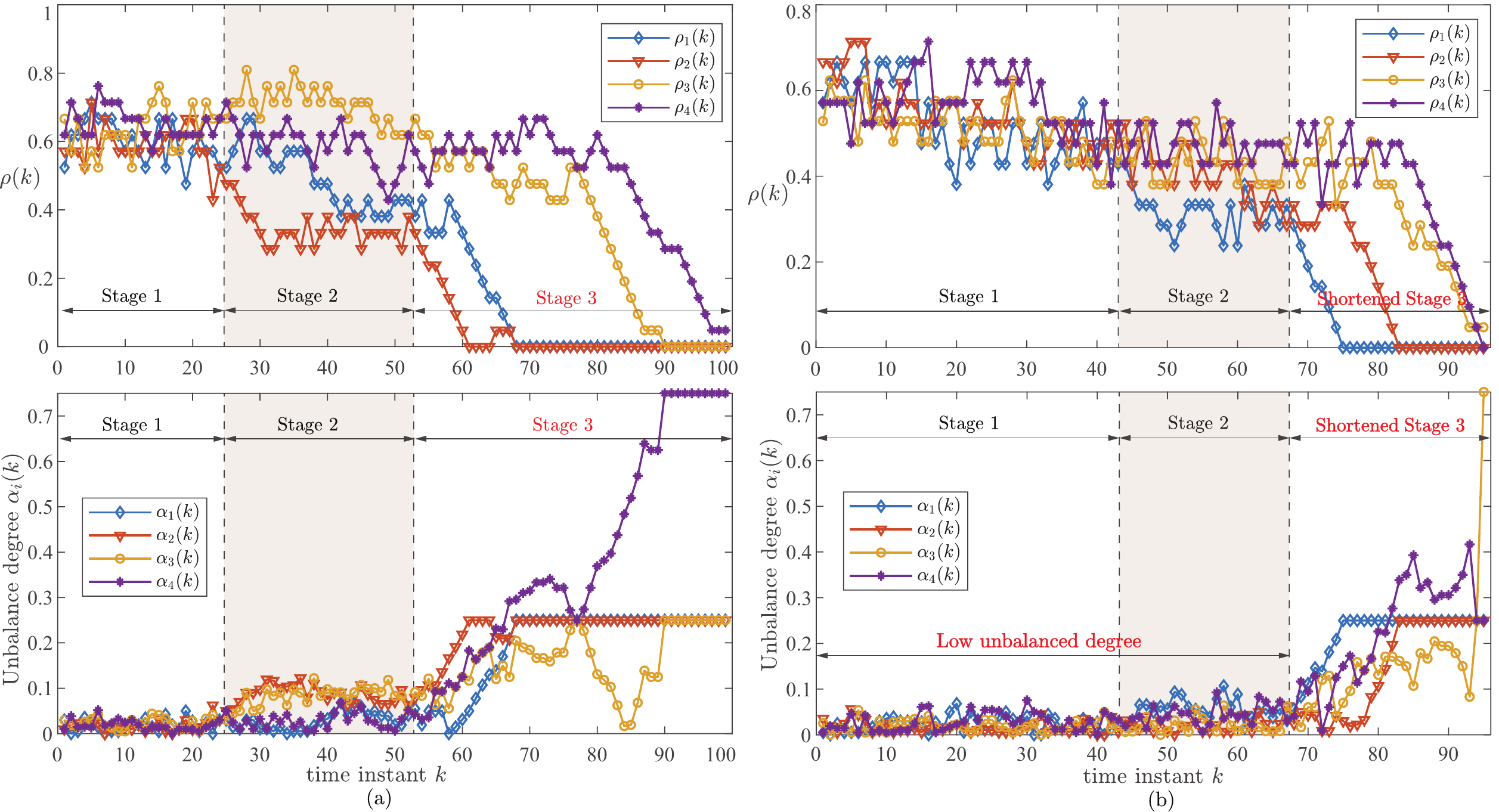}\vspace{-10pt}
  \caption{Trajectories of pedestrian densities and unbalanced degrees. (a) Without control ($u_i(k)=1$, $i\in\mathcal N$, $k=1,2,\ldots,\infty$), (b) Under Bang-bang control \autoref{BB} with $\rho_{\rm aim}=0.5$. The evacuation process without control possesses a long period with extremely unbalanced densities near the 4 exits in (a) (which is referred to as stage 3). In (b), unbalanced behaviors are suppressed under density control and the period of stage 3 is shortened.}\vspace{-8pt}
  \label{fig:6}
\end{figure}

To investigate the effects of density control on evacuation efficiency, we first present the simulation results without the density control, i.e., $u_i(k)=1$, $i\in\mathcal N$, $k=1,2,\ldots,\infty$. In this case, each of the evacuation assistants ignores the real value of pedestrian density near its own exit and tries to attract the pedestrians all the times. Hence, it follows from \autoref{k} that the pedestrians are attracted by the nearest evacuation assistant from their own locations. In the simulation, it is found that the pedestrians used 100 time steps to leave the state space, i.e., $T_{\rm end}=100$. The trajectories of the pedestrian densities near the exits $1,2,3,4$ and the corresponding unbalanced degrees $\alpha_i(k)$, $i\in\mathcal N$, $k=1,2,\ldots,T_{\rm end}$, are shown in \autoref{fig:6} (a).
\begin{figure}
  \centering
  \includegraphics[width=165mm]{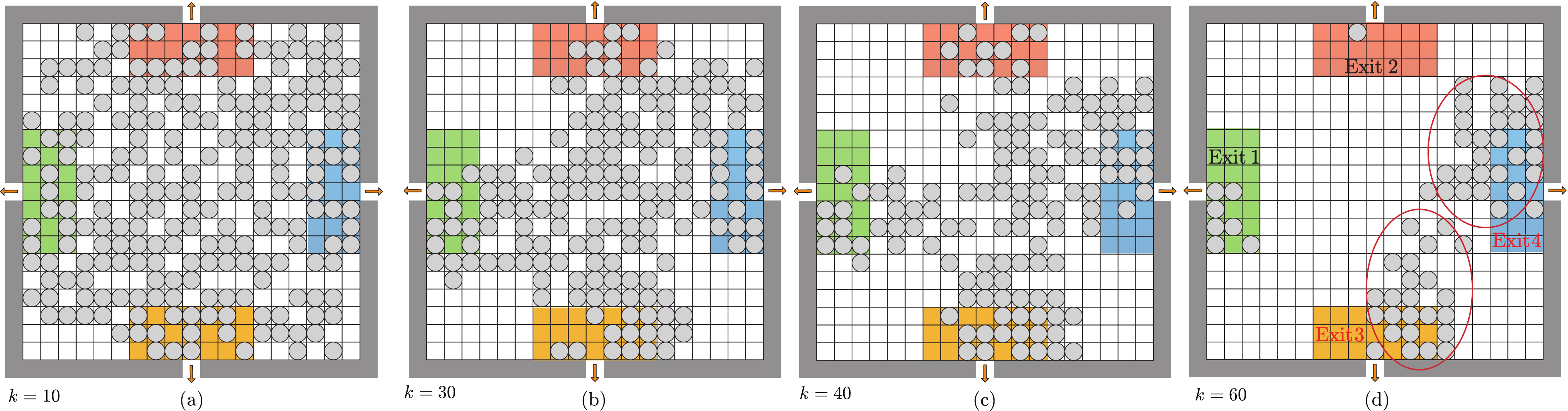}\vspace{-10pt}
  \caption{Pedestrian distributions at the typical time instants without control ($u_i(k)=1$, $i\in\mathcal N$, $k=1,2,\ldots,\infty$). (a): $k=10$, (b): $k=30$, (c): $k=40$, (d): $k=60$. The distributions of pedestrians are more and more unbalanced along with the time instants, especially at $k=60$.}
  \label{fig:7}\vspace{-8pt}
\end{figure}
  It can be seen from the figure that the tendency of pedestrian densities includes 3 obvious different stages. 1) In the beginning (which is henceforth referred to as ``stage 1''), the pedestrian densities in different exits are very similar to each other (e.g., see the typical example of pedestrian distribution at time instant $k=10$ in \autoref{fig:7}(a)); 2) Some slight but visible unbalanced phenomena appear along with the time instants, which we call ``stage 2". The typical examples of pedestrian distribution at stage 2 are illustrated in \refs{fig:7}(b) and \ref{fig:7}(c), where the time instants are given by $k=30$ and $40$, respectively.  It can be seen from those figures that more and more pedestrians begin to be attracted to the exits $3$ and $4$.
  Those slight unbalanced phenomena are caused by the herding effects generated from mutual attractions among the pedestrians. Even if only a few of the pedestrians are closing to one of the exits in an unbalanced way by coincidence, the neighbor pedestrians may be attracted to the same direction and the amount of affected pedestrians may keep growing like a snowballing effect and hence eventually the unbalanced phenomena became extremely visible; 3) Therefore, it is observed from the \emph{later stage} in \autoref{fig:6}(a) that eventually a huge unbalanced behavior appears and the capacities of the exits are not efficiently utilized. The typical example of pedestrian distribution at the ``stage 3'' is shown in \autoref{fig:7}(d) with the time instant given by $k=60$. It can be found from the figure that most of the remaining pedestrians are attracted by the exits $3$ and $4$ but not all the exits.

Recalling that the unbalanced pedestrian densities may bring a larger travel time $T_{\rm end}$ since the capacities of the exits are not efficiently utilized, we begin to present the simulation results
with different density control strategies which adjust the guiding
strength of the evacuation assistants and hence affect the pedestrian densities near the exits in the following sections.

\subsection{Simulation Results with Bang-bang Control}

\subsubsection{An Example Showing Controllable Pedestrian Density}
\begin{figure}
  \centering
  \includegraphics[width=165mm]{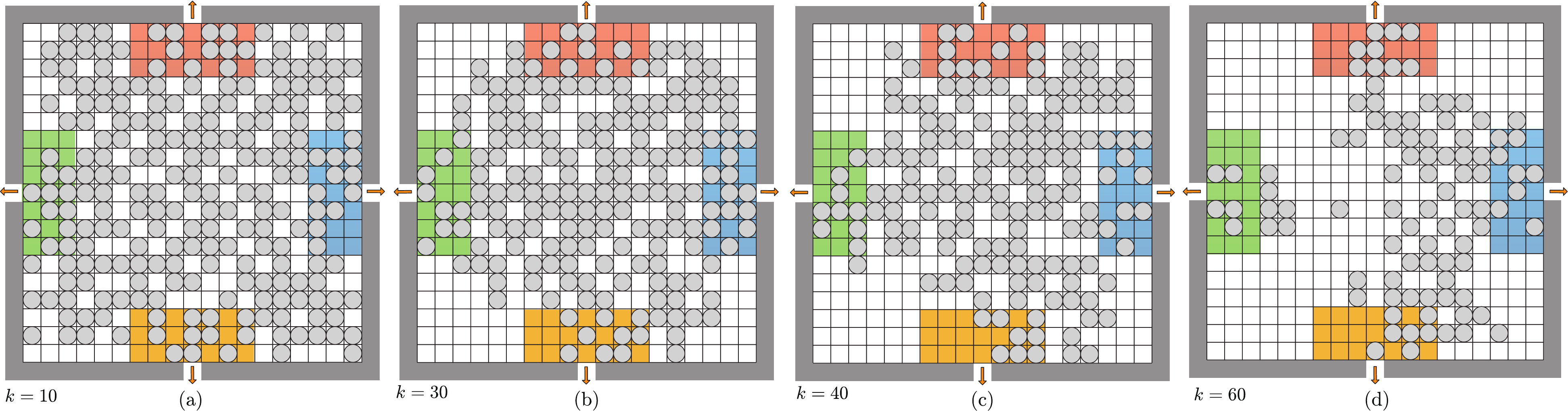}\vspace{-10pt}
  \caption{Pedestrian distributions at the typical time instants under Bang-bang control with $\rho_{\rm aim}=0.5$. (a): $k=10$, (b): $k=30$, (c): $k=40$, (d): $k=60$. Comparing to \autoref{fig:7}, the unbalanced behaviors are certainly suppressed, i.e., distributions of pedestrians are more balanced near the four exits.
  For example, at time instant $k=60$, the crowd is mainly near the exits $2,3$ and $4$ instead of only near the exits $3$ and $4$ in \autoref{fig:7}(d).  }
  \label{fig:8}\vspace{-11pt}
\end{figure}

In this section, we present the simulation results
with Bang-bang control where the guiding coefficients $u_i(k)$, $k=1,2,\ldots,\infty$, are updated by \autoref{BB} for each $i\in\mathcal N$ with the target density to be $\rho_{\rm aim }=0.5$. In this case, each of the evacuation assistants shuts down the guiding signal if the actual pedestrian density is in excess of $\rho_{\rm aim }$ but turns it on for the reverse situation. In the simulation, it is found that the pedestrians used 96 time steps to leave the state space, i.e., $T_{\rm end}=96$, which is slightly shortened comparing to the non-control scenario.
The trajectories of the pedestrian densities near the four exits and the corresponding unbalanced degrees $\alpha_i(k)$, $i\in\mathcal N$, are shown in \autoref{fig:6}(b). The pedestrian distributions at time instants $k=10,30,40$ and $60$ are illustrated in \autoref{fig:8} (same instants as \autoref{fig:7}).
It can be seen from those figures that the herding effects generated from mutual attractions among pedestrians in stage 2 are obviously suppressed under the density control (see \refs{fig:8}(b) and \ref{fig:8}(c)) and hence the pedestrian densities around the four exits possess a low unbalanced degree for a longer period in \autoref{fig:6}(b).
%, even though the remaining amounts of the pedestrians shown in Fig.~\ref{fig:8} are lager than the ones in non-control case in Fig.~\ref{fig:7},
Moreover, since the pedestrian distribution is more balanced, %and hence the capacities of more exits are efficiently utilized,
the period of stage 3 with a huge unbalanced density degree is certainly shortened.

It is worth to note that the above simulation results provide an interesting insight in that the unbalanced density degree is controllable by adjusting the guiding
strength of the evacuation assistants according to the real-time data of pedestrian densities, which may eventually shorten the travel time and hence improve the  efficiency of the evacuation process. To check whether there exists and why there exists an optimized target density, we characterize the influence of the target density $\rho_{\rm aim }$ on pedestrian behaviors under the density control law in \autoref{BB} in the next section.

\vspace{-9pt}
\subsubsection{Influence of the target density in Bang-bang control}\label{BB_reason}
In this section, we note that not all of the density control behaviors could improve evacuation efficiency but a certain set of the target pedestrian density.
The response of target density on travel time of evacuation is shown in \autoref{fig:0} below. It is interesting to see that the travel time $T_{\rm end}$ is shortened under the Bang-bang density control law \autoref{BB} when the target density is moderate (around $0.22<\rho_{\rm aim}<0.6$), and is enlarged when the target is too small ($\rho_{\rm aim}<0.22$).
%For example, the trajectories of pedestrian densities simulated under Bang-bang control with the optimized target density $\rho_{\rm aim}=0.4$ are illustrated in Fig.~\ref{fig:0}(b), where the pedestrian densities are almost synchronized and hence the travel time $T_{\rm end}$ is shortened to $88$ time instants.
We note that the response curve characterized in \autoref{fig:0} is very nature in the sense of physical meaning.
If the target density $\rho_{\rm aim}$ is very large for the Bang-bang density control law in \autoref{BB}, then the pedestrian behaviors are very similar to the ones without density control since it follows from \autoref{BB} that the guiding signals are almost turned on for all the time instants.
Reversely, if the target is very small, then the herding effects among the pedestrians would lead extremely unbalanced distribution during the evacuation process and hence enlarge the travel time, because the guiding signals are shut down at most of the time instants. In this case, the pedestrians are not guided by any evacuation assistant and only move according to the mutual attractions among themselves.
 \begin{figure}
  \centering
  \includegraphics[width=100mm]{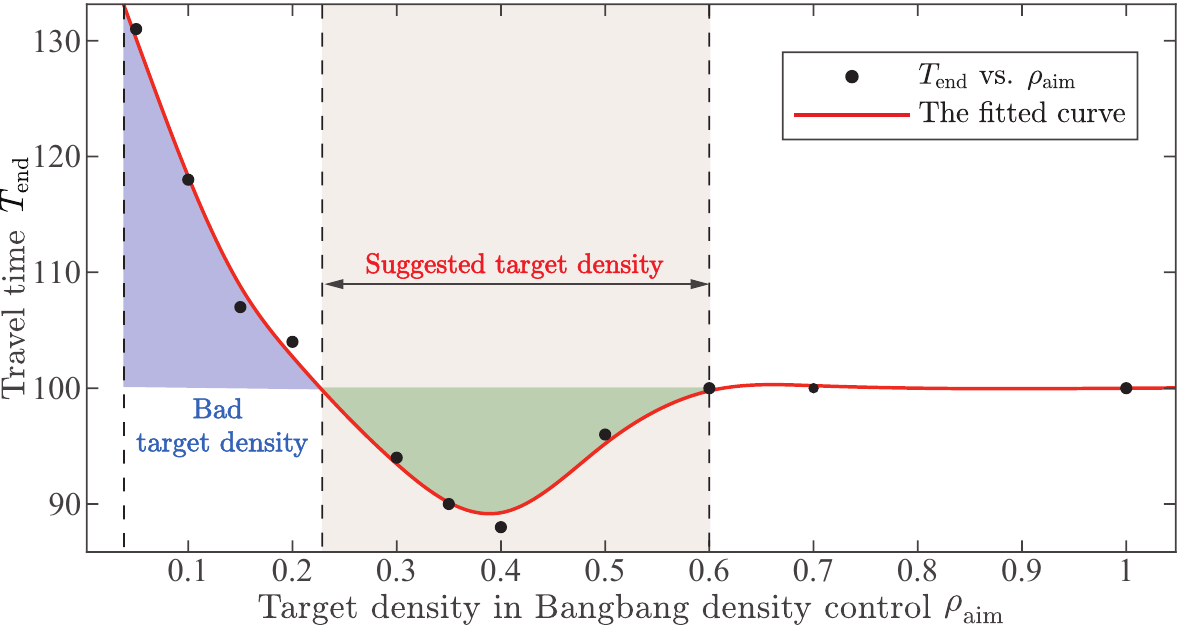}\vspace{-8pt}
  \caption{Response of the target density on travel time under Bang-bang control.   }
  \label{fig:0}\vspace{-8pt}
\end{figure}
\begin{figure}
  \centering
  \includegraphics[width=165mm]{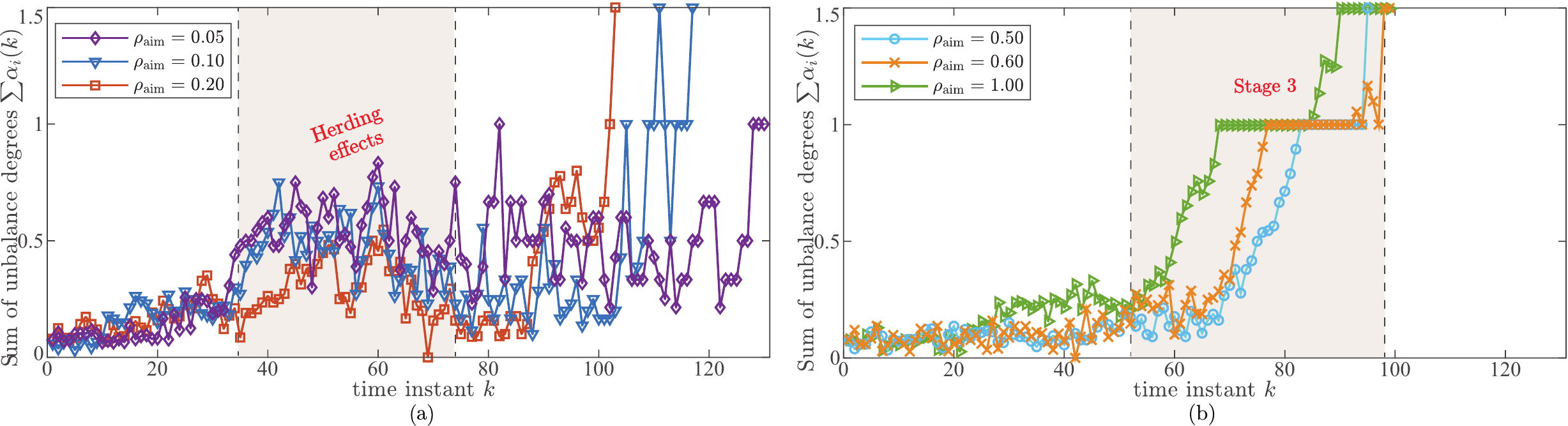}\vspace{-11pt}
  \caption{The influence of the target density on the summation of the unbalanced degrees under Bang-bang control. \textcolor[rgb]{0,0,1}{The herding effects and the stage 3 are suppressed and prolonged in (a) and (b) respectively when the target density increases.} }
  \label{fig:9}\vspace{-10pt}
\end{figure}

\begin{figure}
  \centering
  \includegraphics[width=165mm]{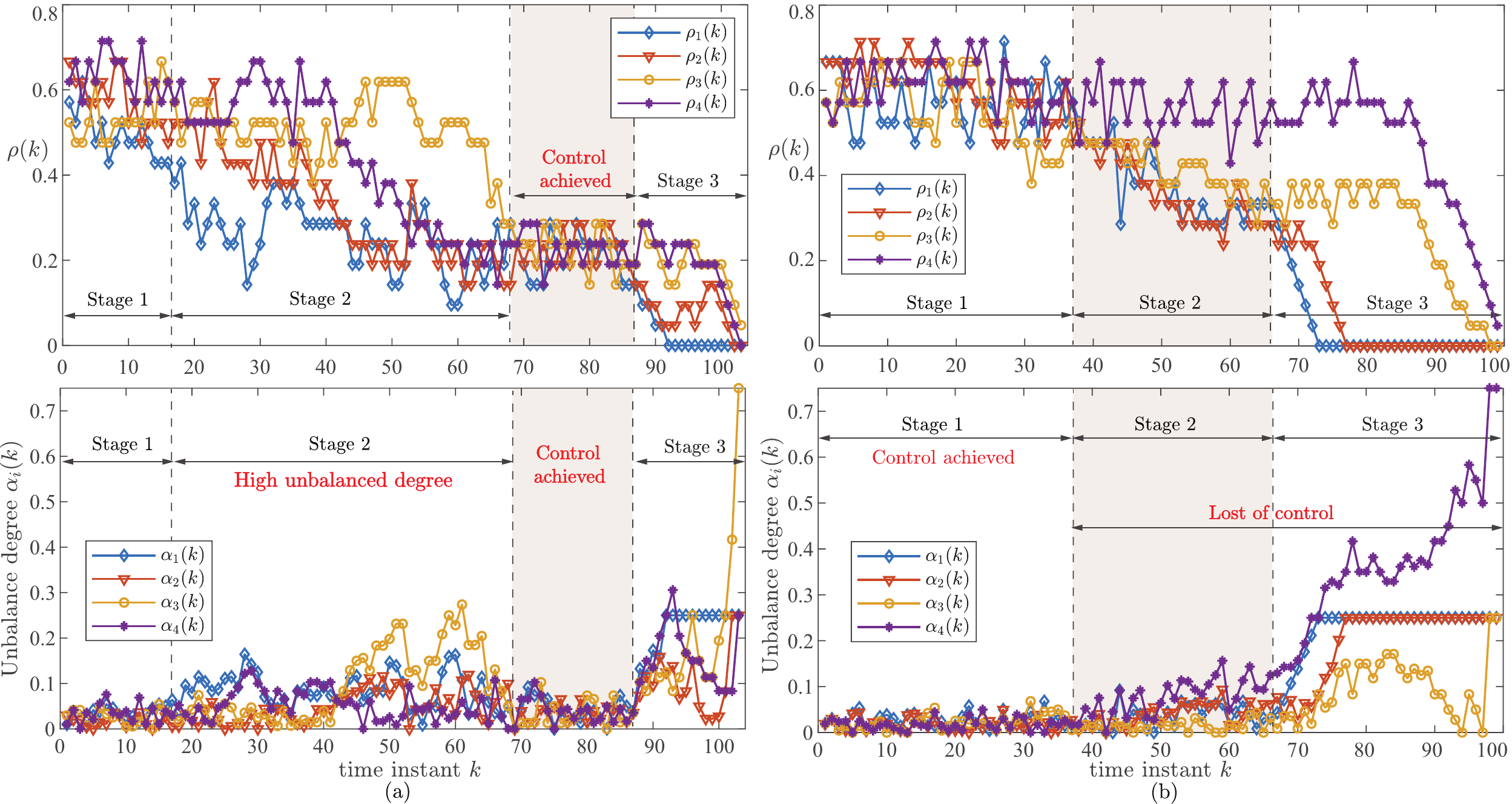}\vspace{-11pt}
  \caption{Trajectories of pedestrian densities and unbalanced degrees under Bang-bang control with \emph{bad} target density $\rho_{\rm aim}$. (a): $\rho_{\rm aim}=0.2$, (b): $\rho_{\rm aim}=0.6$. \textcolor[rgb]{0,0,1}{In (a), the control task is achieved after a while. In (b), the pedestrian densities are initially  controlled well but  {lost of control} soon.} }\vspace{-10pt}
  \label{fig:10}
\end{figure}

To be precise, we verify the above analysis by studying the influence of target density on the unbalanced degrees during the evacuation process.
The trajectories of the summation of the unbalanced degrees with the target densities given by $\rho_{\rm aim}\in\{0.05,0.1,0.2\}$ and $\rho_{\rm aim}\in\{0.5,0.6,1.0\}$
are illustrated in \refs{fig:9}(a) and \ref{fig:9}(b), respectively.
%Here, note thatthe trajectory for $\rho_{\rm aim}=1.0$ is understood the one for the non-control case derived from Fig.~\ref{fig:7}(a).
It can be seen from \autoref{fig:9}(a) that when the target density is very small, the herding effects among pedestrians which lead to unbalanced distributions in stage 2 are extremely huge, but can be suppressed by increasing the target density.
However, \autoref{fig:9}(b) indicates that the Bang-bang control with a too large target density cannot suppress the herding effects well any more and the pedestrian distributions become more unbalanced when the target density increases.
As the typical examples, \autoref{fig:10}(a) shows that the evacuation assistant needs \emph{a long time} to achieve the control task for a too small target density (e.g., $\rho_{\rm aim}=0.2$), and \autoref{fig:10}(b) shows that the pedestrian density is initially under control but \emph{lost of control} soon for a too large target density (e.g., $\rho_{\rm aim}=0.6$).
Consequently, our results demonstrate that to guarantee the well controlled pedestrian density for the entire evacuation process (i.e., to achieve the density control soon and avoid a losing of control), it is required to pick a target density as a middle value (e.g., among $0.2$ and $0.6$).
\vspace{-3pt}
\subsubsection{Shortness of Bang-bang Control: Sacrificing the efficiency during the early stage}\label{dis_BB}
On the one hand, although the Bang-bang density control scheme in \autoref{BB} with a properly designed target density indeed improves the evacuation efficiency, it is surprised to observe that the pedestrians remaining in the state space under Bang-bang control are more than the case without density control for each of the time instants shown in \refs{fig:7} and \ref{fig:8}. Hence, we conclude that the density balance contributing to shorten the total travel time by Bang-bang control is yielded
from sacrificing the efficiency during the early stage of the evacuation process (see the solid proof from simulation in \autoref{PI_c} below). Recalling that the fundamental factor affecting the evacuation efficiency is the efficient and balanced utilization of all the exits, the Bang-bang control scheme in \autoref{BB} maybe not a best choice for the guiding strength update laws to improve the evacuation efficiency. It is natural to ask that is there any other control scheme to guarantee not only the balanced pedestrian densities but also the efficient utilization of the exits.
On the other hand, we note that the Bang-bang control is pretty \emph{sensitive} with respect to the observed (current) density value, which may be the reason of
efficiency loss in the evacuation process.
Hence, we present the results simulated by a different control scheme in the following section.
\vspace{-7pt}
\subsection{Simulation Results with PI Control}\label{PI_c}
In this section, we present the simulation results
under the PI control scheme with $K_p=70$ and $K_i=20$, where the guiding coefficients $u_i(k)$, $k=1,2,\ldots,\infty$, are updated by \autoref{8} for each $i\in\mathcal N$. In this case, instead of using an on-off guiding signal, each of the evacuation assistants may slightly adjust the guiding coefficient from 0 to 1 when there is a density error between the real-time density and the target density.
\begin{figure}\vspace{-8pt}
  \centering
  \includegraphics[width=100mm]{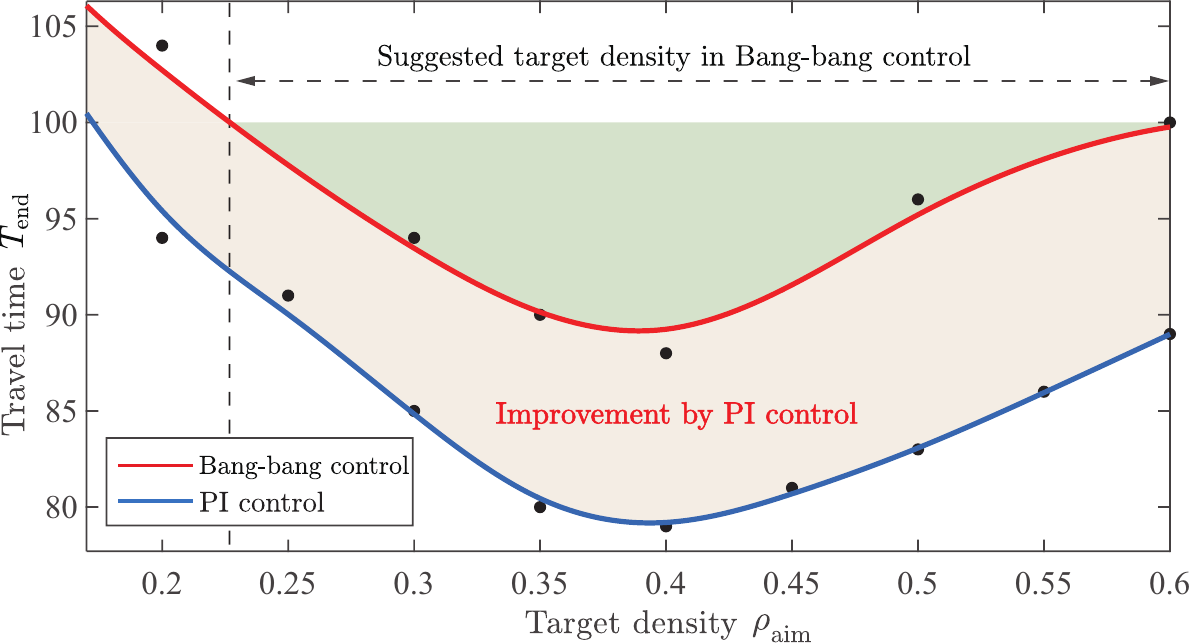}\vspace{-10pt}
  \caption{Comparison of travel time response with respect to target densities for PI control \textcolor[rgb]{0,0,1}{in} \autoref{8} and Bang-bang control \textcolor[rgb]{0,0,1}{in} \autoref{BB}. The evacuation efficiency is improved by PI control for the suggested target density of the Bang-bang control.}
  \label{fig:22}
\end{figure}

 The response of target density on travel time of evacuation is shown in \autoref{fig:22}. It can be see from the figure that the tendency of the response curve generated by PI control is very similar to the one by Bang-bang control, but the evacuation efficiency is significantly improved.
The reason why a too large or too small target density brings a bigger travel time is already interpreted in \autoref{BB_reason}, that is, a too large target density leads to a \emph{loss of control} soon and a too small target density leads \emph{a long time} for achieving the control task.

\begin{figure}
  \centering
  \includegraphics[width=165mm]{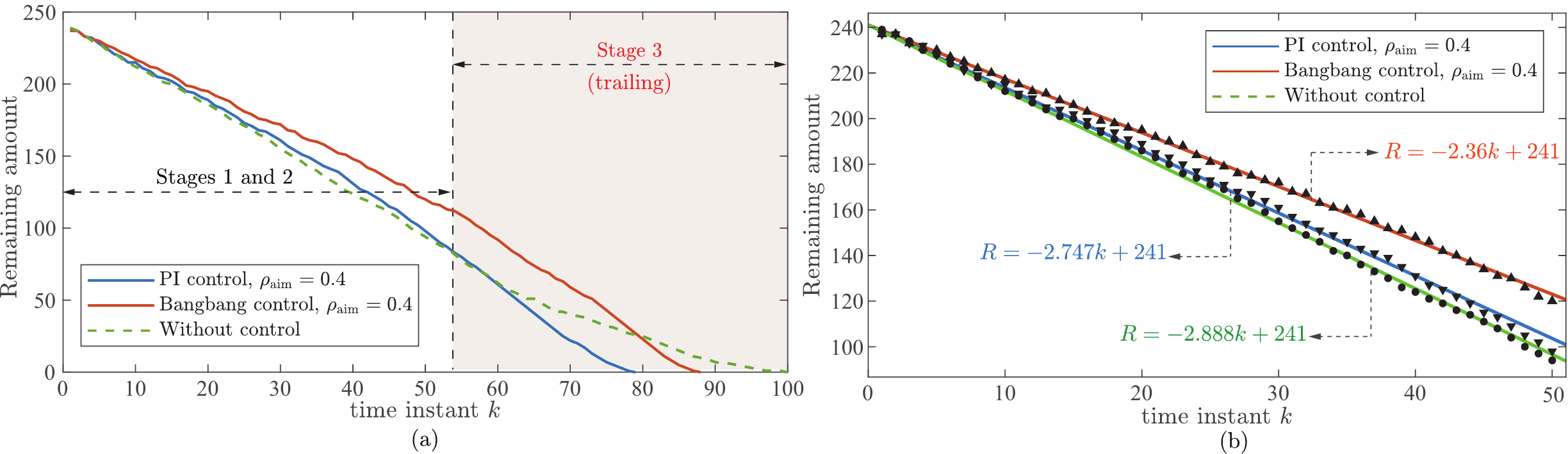}\vspace{-10pt}
  \caption{Amount of remaining pedestrians in the state space under cases of non-control, Bang-bang control and PI control with the optimized target density. Note that (b) is the copy of (a) with fitted straight lines, which shows that in the first 50 time instants, around 2.888, 2.747 and 2.36 pedestrians left the state space per each time instant.}
  \label{fig:13}\vspace{-4pt}
\end{figure}
\begin{figure}
  \centering
  \includegraphics[width=165mm]{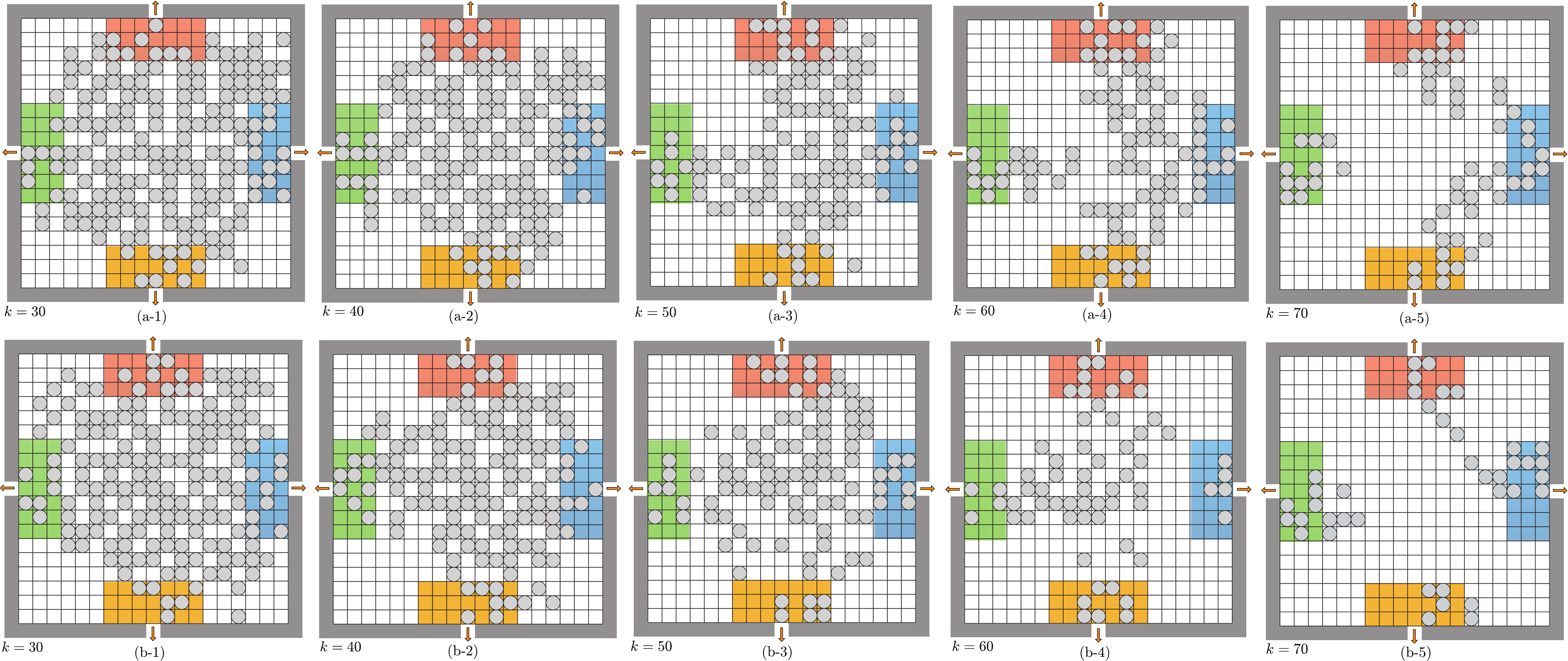}\vspace{-10pt}
  \caption{Pedestrian distributions at the typical time instants under Bang-bang and PI control with $\rho_{\rm aim}=0.4$. (a): Bang-bang, (b): PI. The crowd in the middle in (a) keeps moving to the right side and the tendency is not suppressed by the Bang-bang control in (a-3)--(a-5). But the tendency is keeping corrected to the reverse directions by the PI control in (b-3)--(b-5). That is to say, the density control for the \emph{partial} regions (near the exits) could yield a better \emph{global effect} for balancing the pedestrians in the \emph{rest of the regions} if we use the PI control than Bang-bang control.}
  \label{fig:11}\vspace{-10pt}
\end{figure}
\begin{figure}
  \centering
  \includegraphics[width=165mm]{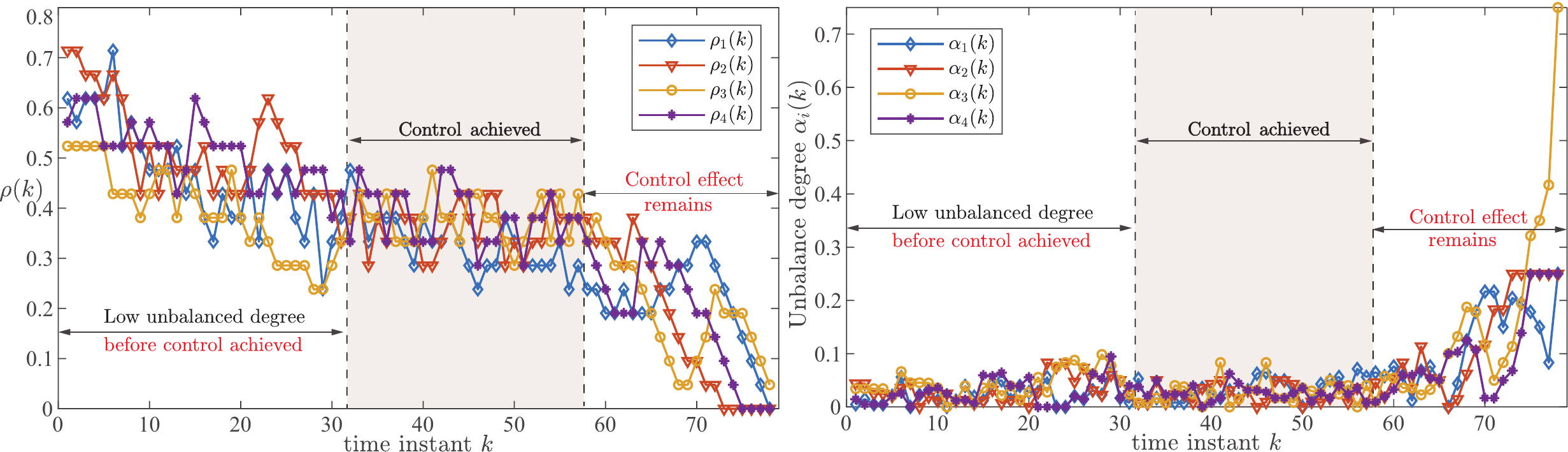}\vspace{-10pt}
  \caption{Trajectories of pedestrian densities and unbalanced degrees under designed density control scheme (PI control with $\rho_{\rm aim}=0.4$). The control target is well achieved comparing to the curves shown in Figs.~\ref{fig:7} and \ref{fig:10}.}
  \label{fig:12}\vspace{-4pt}
\end{figure}

Now, we study the principle of why PI control improves the evacuation efficiency by making comparisons with the cases of non-control and Bang-bang control.
The amounts of remaining pedestrians simulated for the PI control and Bang-bang control under the best target density $\rho_{\rm aim}=0.4$ are shown in \autoref{fig:13} along with the trajectory for the non-control case. All of the trajectories possesses an almost liner tendency in the early stage of the evacuation process.
But the trajectory for the non-control case possesses an obvious ``tail" (caused by the unbalanced pedestrian distribution) in the later stage (which is referred to as stage 3 in \autoref{without}) during which the evacuation efficiency declines significantly.
  Note that the corresponding trajectories of the density at stage 3 are already given in \autoref{fig:7}(a) above.
  After we apply the Bang-bang control, the ``obvious tail" is indeed eliminated and hence the total travel time is shortened.
  Moreover, \autoref{fig:13}(a) further verifies our analysis given in \autoref{dis_BB} in the sense that the illustrated curves for Bang-bang control is much gentler than the one without control. In particular, it follows from the fitted lines of the first 50 time instants shown in \autoref{fig:7}(b) that
  around 2.888
pedestrians left the state space per each time instant for the non-control case, but only 2.36 pedestrians left under Bang-bang control, which indicates that the elimination of the ``obvious tail" via Bang-bang control is made by sacrificing the efficiency during the \emph{early stage}.
However, for the PI control case, it can be seen from \autoref{fig:13}(a) that the ``obvious tail" is eliminated without losing too much efficiency.
Besides, the pedestrian distributions shown in \autoref{fig:11} demonstrate that the PI control possesses a better global performance for balancing the pedestrian distribution in the entire state space than the Bang-bang control, because the crowd in \autoref{fig:11}(a) keeps moving towards the right-hand side in an unbalanced way without any obvious suppressions, but the unbalanced tendency in \autoref{fig:11}(b) keeps being corrected to the reverse directions (see \autoref{fig:11}(b-3)--(b-5)). For the convenience of the reader to make a comparison with the trajectories shown in \refs{fig:6} and \ref{fig:10}, we illustrate the trajectories of pedestrian densities and unbalanced degrees under PI control with the optimized target density in \autoref{fig:12}, which shows an almost synchronization behavior during the entire evacuation process.  The above results give an important insight into designing the guiding strategies in the real evacuation process.

\section{Conclusion}\label{section:4}
%Different from the existing works in terms of guided emergency evacuation which only consider a  single exit evacuation scenario based on a social force model.
%For example, Yang et al. \cite{yang2016necessity} first claimed the necessity of guides in pedestrian  .
%Ma et al. \cite{ma2016effective,ma2017dual} considered dynamic leaders inside a room who attract the crowds and move together with them towards a unique exit with different  pedestrian densities.
%The possibility and effects of controlling the pedestrian densities were not considered yet in the literature.
{To reveal the possibility of enhancing guided evacuation efficiency by imposing control schemes, in this paper,
we investigated the effect of variable guiding strategies in a multi-exits evacuation process.
Specifically}, we proposed a force-driven cellular automaton model with adjustable guiding attractions imposed by the evacuation assistants located in the exits.
In this model, each of the evacuation assistants tries to attract the pedestrians towards its own exit by sending a quantifiable guiding signal, which may be adjusted according to the values of pedestrian density near the exit.
In accordance to the cases with and without density control, the simulation results were respectively established.

According to the evidence shown in the simulation, we first revealed the fact that an inherent problem of \emph{unbalanced} pedestrian distribution exists in the multi-exits evacuation process, i.e., an extremely unbalanced pedestrian distribution exists in the later stage of the evacuation process, which eventually leads to inefficient utilization of the exits.  Our results showed that the \emph{unbalanced} pedestrian distribution is mainly yielded by a snowballing effect generated from the mutual attractions among the pedestrians, and can be suppressed by partially controlling the pedestrian densities around the exits.
We considered the effects of the density controls in an on-off fashion (i.e., Bang-bang control) and a quantifiable fashion (i.e., PI control) to adjust the guiding strength for each of the evacuation assistants.
A moderate value representing the optimized target density for both of the Bang-bang control and PI control was found.
It was revealed that the efficiency improvement of Bang-bang control is based on
sacrificing the efficiency during the early stage, but the one for PI control is yielded without losing too much efficiency during the early stage and hence PI control further improves the evacuation efficiency. Since we showed that the density control for the partial regions (near the exits)
could yield a global effect for balancing the pedestrians in the rest of the regions, our results are expected to give some interesting and important insights to design the guiding strategies in the real evacuation process.

{However, the control performance of Bang-bang control and PI control might be strongly connected to the morphic conditions of the evacuation space including the number of exits \cite{kurdi2018effect}, asymmetric degree of exits \cite{yue2011simulation2} and sampling regions. For yielding more general result, the analysis on those factors may be promised necessary future research directions.}
The other future directions may include the extensions to multi-exits coordination and the game formulations between different clusters \cite{tanimoto2015fundamentals}.
\section*{Acknowledgments}
This work was supported jointly by Program for Young Excellent Talents in University of Fujian Province (201847) and China Scholarship Council (201908050058).



%% References


\bibliographystyle{elsarticle-num}
\bibliography{revised2}

\begin{thebibliography}{10}
\expandafter\ifx\csname url\endcsname\relax
  \def\url#1{\texttt{#1}}\fi
\expandafter\ifx\csname urlprefix\endcsname\relax\def\urlprefix{URL }\fi
\expandafter\ifx\csname href\endcsname\relax
  \def\href#1#2{#2} \def\path#1{#1}\fi

\bibitem{meng2019pedestrian}
Q.~Meng, M.~Zhou, J.~Liu, H.~Dong, Pedestrian evacuation with herding behavior
  in the view-limited condition, IEEE Trans. Comput. Soc. Syst. 6~(3) (2019)
  567--575.
\newblock \href {http://dx.doi.org/https://doi.org/10.1109/TCSS.2019.2915772}
  {\path{doi:https://doi.org/10.1109/TCSS.2019.2915772}}.

\bibitem{cheng2018emergence}
Y.~Cheng, X.~Zheng, Emergence of cooperation during an emergency evacuation,
  Appl. Math. Comput. 320 (2018) 485--494.
\newblock \href {http://dx.doi.org/https://doi.org/10.1016/j.amc.2017.10.011}
  {\path{doi:https://doi.org/10.1016/j.amc.2017.10.011}}.

\bibitem{haghani2019panic}
M.~Haghani, E.~Cristiani, N.~W. Bode, M.~Boltes, A.~Corbetta, Panic,
  irrationality, and herding: three ambiguous terms in crowd dynamics research,
  J. Adv. Transp. 2019.
\newblock \href {http://dx.doi.org/https://doi.org/10.1155/2019/9267643}
  {\path{doi:https://doi.org/10.1155/2019/9267643}}.

\bibitem{guan2019towards}
J.~Guan, K.~Wang, Towards pedestrian room evacuation with a spatial game, Appl.
  Math. Comput. 347 (2019) 492--501.
\newblock \href {http://dx.doi.org/https://doi.org/10.1016/j.amc.2018.11.003}
  {\path{doi:https://doi.org/10.1016/j.amc.2018.11.003}}.

\bibitem{lovaas1994modeling}
G.~G. L{\o}v{\aa}s, Modeling and simulation of pedestrian traffic flow, Transp.
  Res. Part B 28~(6) (1994) 429--443.
\newblock \href
  {http://dx.doi.org/https://doi.org/10.1016/0191-2615(94)90013-2}
  {\path{doi:https://doi.org/10.1016/0191-2615(94)90013-2}}.

\bibitem{helbing2000simulating}
D.~Helbing, I.~Farkas, T.~Vicsek, Simulating dynamical features of escape
  panic, Nature 407~(6803) (2000) 487--490.
\newblock \href {http://dx.doi.org/https://doi.org/10.1038/35035023}
  {\path{doi:https://doi.org/10.1038/35035023}}.

\bibitem{muller2014study}
F.~M{\"u}ller, O.~Wohak, A.~Schadschneider, Study of influence of groups on
  evacuation dynamics using a cellular automaton model, Transp. Res. Rec. 2
  (2014) 168--176.
\newblock \href {http://dx.doi.org/https://doi.org/10.1016/j.trpro.2014.09.022}
  {\path{doi:https://doi.org/10.1016/j.trpro.2014.09.022}}.

\bibitem{PhysRevE}
D.~Yanagisawa, K.~Nishinari, Mean-field theory for pedestrian outflow through
  an exit, Phys. Rev. E 76 (2007) 061117.
\newblock \href {http://dx.doi.org/https://doi.org/10.1103/PhysRevE.76.061117}
  {\path{doi:https://doi.org/10.1103/PhysRevE.76.061117}}.

\bibitem{varas2007cellular}
A.~Varas, M.~Cornejo, D.~Mainemer, B.~Toledo, J.~Rogan, V.~Munoz, J.~Valdivia,
  Cellular automaton model for evacuation process with obstacles, Physica A
  382~(2) (2007) 631--642.
\newblock \href {http://dx.doi.org/https://doi.org/10.1016/j.physa.2007.04.006}
  {\path{doi:https://doi.org/10.1016/j.physa.2007.04.006}}.

\bibitem{huang2017behavior}
K.~Huang, X.~Zheng, Y.~Cheng, Y.~Yang, Behavior-based cellular automaton model
  for pedestrian dynamics, Appl. Math. Comput. 292 (2017) 417--424.
\newblock \href {http://dx.doi.org/https://doi.org/10.1016/j.amc.2016.07.002}
  {\path{doi:https://doi.org/10.1016/j.amc.2016.07.002}}.

\bibitem{alizadeh2011dynamic}
R.~Alizadeh, A dynamic cellular automaton model for evacuation process with
  obstacles, Saf. Sci. 49~(2) (2011) 315--323.
\newblock \href {http://dx.doi.org/https://doi.org/10.1016/j.ssci.2010.09.006}
  {\path{doi:https://doi.org/10.1016/j.ssci.2010.09.006}}.

\bibitem{cao2018exit}
S.~Cao, L.~Fu, W.~Song, Exit selection and pedestrian movement in a room with
  two exits under fire emergency, Appl. Math. Comput. 332 (2018) 136--147.
\newblock \href {http://dx.doi.org/https://doi.org/10.1016/j.amc.2018.03.048}
  {\path{doi:https://doi.org/10.1016/j.amc.2018.03.048}}.

\bibitem{shahhoseini2019pedestrian}
Z.~Shahhoseini, M.~Sarvi, Pedestrian crowd flows in shared spaces:
  investigating the impact of geometry based on micro and macro scale measures,
  Transp. Res. Part B 122 (2019) 57--87.
\newblock \href {http://dx.doi.org/https://doi.org/10.1016/j.trb.2019.01.019}
  {\path{doi:https://doi.org/10.1016/j.trb.2019.01.019}}.

\bibitem{huang2015behavioral}
K.~Huang, X.~Zheng, Y.~Yang, T.~Wang, Behavioral evolution in evacuation crowd
  based on heterogeneous rationality of small groups, Appl. Math. Comput. 266
  (2015) 501--506.
\newblock \href {http://dx.doi.org/https://doi.org/10.1016/j.amc.2015.05.065}
  {\path{doi:https://doi.org/10.1016/j.amc.2015.05.065}}.

\bibitem{chraibi2010generalized}
M.~Chraibi, A.~Seyfried, A.~Schadschneider, Generalized centrifugal-force model
  for pedestrian dynamics, Phys. Rev. E 82~(4) (2010) 046111.
\newblock \href {http://dx.doi.org/https://doi.org/10.1103/PhysRevE.82.046111}
  {\path{doi:https://doi.org/10.1103/PhysRevE.82.046111}}.

\bibitem{liao2014layout}
W.~Liao, X.~Zheng, L.~Cheng, Y.~Zhao, Y.~Cheng, Y.~Wang, Layout effects of
  multi-exit ticket-inspectors on pedestrian evacuation, Saf. Sci. 70 (2014)
  1--8.
\newblock \href {http://dx.doi.org/https://doi.org/10.1016/j.ssci.2014.04.015}
  {\path{doi:https://doi.org/10.1016/j.ssci.2014.04.015}}.

\bibitem{hao2014exit}
Y.~Hao, Z.~Bin-Ya, S.~Chun-Fu, X.~Yan, Exit selection strategy in pedestrian
  evacuation simulation with multi-exits, Chin. Phys. B 23~(5) (2014) 050512.
\newblock \href
  {http://dx.doi.org/https://doi.org/10.1088/1674-1056/23/5/050512}
  {\path{doi:https://doi.org/10.1088/1674-1056/23/5/050512}}.

\bibitem{HRABAK2017486}
P.~Hrabak, M.~Bukacek, Influence of agents heterogeneity in cellular model of
  evacuation, J. Comput. Sci. 21 (2017) 486 -- 493.
\newblock \href {http://dx.doi.org/https://doi.org/10.1016/j.jocs.2016.08.002}
  {\path{doi:https://doi.org/10.1016/j.jocs.2016.08.002}}.

\bibitem{chen2012study}
C.~K. Chen, J.~Li, D.~Zhang, Study on evacuation behaviors at a {T}-shaped
  intersection by a force-driving cellular automata model, Physica A 391~(7)
  (2012) 2408--2420.
\newblock \href {http://dx.doi.org/https://doi.org/10.1016/j.physa.2011.12.001}
  {\path{doi:https://doi.org/10.1016/j.physa.2011.12.001}}.

\bibitem{li2019extended}
X.~Li, F.~Guo, H.~Kuang, Z.~Geng, Y.~Fan, An extended cost potential field
  cellular automaton model for pedestrian evacuation considering the
  restriction of visual field, Physica A 515 (2019) 47--56.
\newblock \href {http://dx.doi.org/https://doi.org/10.1016/j.physa.2018.09.145}
  {\path{doi:https://doi.org/10.1016/j.physa.2018.09.145}}.

\bibitem{huang2017weighted}
K.~Huang, X.~Zheng, A weighted evolving network model for pedestrian
  evacuation, Appl. Math. Comput. 298 (2017) 57--64.
\newblock \href {http://dx.doi.org/https://doi.org/10.1016/j.amc.2016.10.040}
  {\path{doi:https://doi.org/10.1016/j.amc.2016.10.040}}.

\bibitem{pereira2017emergency}
L.~A. Pereira, D.~Burgarelli, L.~Duczmal, F.~Cruz, Emergency evacuation models
  based on cellular automata with route changes and group fields, Physica A 473
  (2017) 97--110.
\newblock \href {http://dx.doi.org/https://doi.org/10.1016/j.physa.2017.01.048}
  {\path{doi:https://doi.org/10.1016/j.physa.2017.01.048}}.

\bibitem{miyagawa2020cellular}
D.~Miyagawa, G.~Ichinose, Cellular automaton model with turning behavior in
  crowd evacuation, Physica A: Statistical Mechanics and its Applications
  (2020) 124376\href
  {http://dx.doi.org/https://doi.org/10.1016/j.physa.2020.124376}
  {\path{doi:https://doi.org/10.1016/j.physa.2020.124376}}.

\bibitem{yang2016necessity}
X.~Yang, H.~Dong, X.~Yao, X.~Sun, Q.~Wang, M.~Zhou, Necessity of guides in
  pedestrian emergency evacuation, Physica A 442 (2016) 397--408.
\newblock \href {http://dx.doi.org/https://doi.org/10.1016/j.physa.2015.08.020}
  {\path{doi:https://doi.org/10.1016/j.physa.2015.08.020}}.

\bibitem{ma2016effective}
Y.~Ma, R.~K.~K. Yuen, E.~W.~M. Lee, Effective leadership for crowd evacuation,
  Physica A 450 (2016) 333--341.
\newblock \href {http://dx.doi.org/https://doi.org/10.1016/j.physa.2015.12.103}
  {\path{doi:https://doi.org/10.1016/j.physa.2015.12.103}}.

\bibitem{ma2017dual}
Y.~Ma, E.~W.~M. Lee, M.~Shi, Dual effects of guide-based guidance on pedestrian
  evacuation, Mod. Phys. Lett. 381~(22) (2017) 1837--1844.
\newblock \href
  {http://dx.doi.org/https://doi.org/10.1016/j.physleta.2017.03.050}
  {\path{doi:https://doi.org/10.1016/j.physleta.2017.03.050}}.

\bibitem{zhou2019guided}
M.~Zhou, H.~Dong, P.~A. Ioannou, Y.~Zhao, F.-Y. Wang, Guided crowd evacuation:
  approaches and challenges, IEEE/CAA Journal of Automatica Sinica 6~(5) (2019)
  1081--1094.
\newblock \href {http://dx.doi.org/https://doi.org/10.1109/JAS.2019.1911672}
  {\path{doi:https://doi.org/10.1109/JAS.2019.1911672}}.

\bibitem{zhou2019optimization}
M.~Zhou, H.~Dong, Y.~Zhao, P.~A. Ioannou, F.-Y. Wang, Optimization of crowd
  evacuation with leaders in urban rail transit stations, IEEE Trans. Intell.
  Transp. Syst. 20~(12) (2019) 4476--4487.
\newblock \href {http://dx.doi.org/10.1109/TITS.2018.2886415}
  {\path{doi:10.1109/TITS.2018.2886415}}.

\bibitem{long2020simulation}
S.~Long, D.~Zhang, S.~Li, S.~Yang, B.~Zhang, Simulation-based model of
  emergency evacuation guidance in the metro stations of {C}hina, IEEE Access 8
  (2020) 62670--62688.
\newblock \href {http://dx.doi.org/https://doi.org/10.1109/ACCESS.2020.2983441}
  {\path{doi:https://doi.org/10.1109/ACCESS.2020.2983441}}.

\bibitem{yang2020guide}
X.~Yang, X.~Yang, Q.~Wang, Y.~Kang, F.~Pan, Guide optimization in pedestrian
  emergency evacuation, Appl. Math. Comput. 365 (2020) 124711.
\newblock \href {http://dx.doi.org/https://doi.org/10.1016/j.amc.2019.124711}
  {\path{doi:https://doi.org/10.1016/j.amc.2019.124711}}.

\bibitem{kurdi2020balanced}
H.~Kurdi, A.~Almulifi, S.~Al-Megren, K.~Youcef-Toumi, A balanced evacuation
  algorithm for facilities with multiple exits, Eur. J. Oper. Res. 289 (2021)
  285--296.
\newblock \href {http://dx.doi.org/https://doi.org/10.1016/j.ejor.2020.07.012}
  {\path{doi:https://doi.org/10.1016/j.ejor.2020.07.012}}.

\bibitem{tanimoto2010study}
J.~Tanimoto, A.~Hagishima, Y.~Tanaka, Study of bottleneck effect at an
  emergency evacuation exit using cellular automata model, mean field
  approximation analysis, and game theory, Physica A 389~(24) (2010)
  5611--5618.
\newblock \href {http://dx.doi.org/https://doi.org/10.1016/j.physa.2010.08.032}
  {\path{doi:https://doi.org/10.1016/j.physa.2010.08.032}}.

\bibitem{tanimoto2019evolutionary}
J.~Tanimoto, Evolutionary games with sociophysics: Analysis of traffic flow and
  epidemics, Springer, 2019.
\newblock \href {http://dx.doi.org/https://doi.org/10.1007/978-981-13-2769-8}
  {\path{doi:https://doi.org/10.1007/978-981-13-2769-8}}.

\bibitem{kurdi2018effect}
H.~A. Kurdi, S.~Al-Megren, R.~Althunyan, A.~Almulifi, Effect of exit placement
  on evacuation plans, Eur. J. Oper. Res. 269~(2) (2018) 749--759.
\newblock \href {http://dx.doi.org/https://doi.org/10.1016/j.ejor.2018.01.050}
  {\path{doi:https://doi.org/10.1016/j.ejor.2018.01.050}}.

\bibitem{yue2011simulation2}
H.~Yue, H.~Guan, C.~Shao, X.~Zhang, Simulation of pedestrian evacuation with
  asymmetrical exits layout, Physica A 390~(2) (2011) 198--207.
\newblock \href {http://dx.doi.org/https://doi.org/10.1016/j.physa.2010.10.003}
  {\path{doi:https://doi.org/10.1016/j.physa.2010.10.003}}.

\bibitem{tanimoto2015fundamentals}
J.~Tanimoto, Fundamentals of evolutionary game theory and its applications,
  Springer, 2015.
\newblock \href {http://dx.doi.org/https://doi.org/10.1007/978-4-431-54962-8}
  {\path{doi:https://doi.org/10.1007/978-4-431-54962-8}}.

\end{thebibliography}






% \end{thebibliography}

\end{document}